\documentclass[aps,onecolumn,nopacs,nofootinbib,floatfix,superscriptaddress]{revtex4}
\usepackage{graphicx}
\usepackage{amsfonts}
\usepackage{amssymb}
\usepackage{amsbsy}
\usepackage{hyperref}
\usepackage{amsmath}
\usepackage{mathrsfs}
\usepackage{latexsym}
\usepackage{natbib}
\usepackage{bm}
\usepackage{subfigure} 
\usepackage{color}
\usepackage{wasysym}
\usepackage{mathbbol}
\usepackage{bigints}
\allowdisplaybreaks
\usepackage[normalem]{ulem}
\usepackage[dvipsnames]{xcolor}
\usepackage{multirow}
\usepackage{braket}
\usepackage[normalem]{ulem}

\definecolor{napiergreen}{rgb}{0.16, 0.5, 0.0}

\renewcommand*{\l}{\lambda_{\star}}

\def\l{\left}
\def\r{\right}
\def\DM{{\rm d}}

\begin{document}

\title{Twin-paradox and Entanglement}

\author{K. Hari}
\email{hari.k@iitb.ac.in}
\affiliation{Department of Physics, Indian Institute of Technology Bombay, Mumbai 400076, India}

\author{Subhajit Barman}
\email{subhajit.barman@physics.iitm.ac.in}
\affiliation{Centre for Strings, Gravitation and Cosmology, Department of Physics, Indian Institute of Technology Madras, Chennai 600036, India}

\author{Dawood Kothawala}
\email{dawood@iitm.ac.in}
\affiliation{Centre for Strings, Gravitation and Cosmology, Department of Physics, Indian Institute of Technology Madras, Chennai 600036, India}

\begin{abstract}
\noindent We study the quantum version of the classical twin paradox in special relativity by replacing the twins with quantum detectors, and studying the transitions and entanglement induced by coupling them to a quantum field. We show that the \textit{changes} in direction of acceleration leave imprints on detector responses and entanglement, inducing novel features which might have relevance in black hole spacetimes.

\end{abstract}

\date{\today}

\maketitle

\section{Introduction}\label{sec:Introduction}

The twin ``paradox" is perhaps one of the simplest results in special relativity that demonstrates, very clearly, the operational significance of proper time intervals in physical systems. In its classical form, the paradox is most clearly stated by explicitly computing differential ageing of twins on smooth trajectories, one of which is inertial while the other one accelerated. A rather un-physical choice for the latter is the Rindler trajectory, which nevertheless gives results which capture the most basic features of the paradox. A more physically realistic trajectories give similar results \cite{Minguzzi:2004fa}. 

In the quantum regime, differential ageing or time dilation between two physical systems acquires a more interesting aspect: it can be used to study genuinely quantum features such as entanglement and decoherence. Indeed, entanglement dynamics from the perspective of differential ageing has been studied in simple settings such as pair of inertial and uniformly accelerating detectors. It was shown in Ref. \cite{Fuentes-Schuller:2004iaz} that the entanglement between inertial and uniformly accelerated observer as measured using negativity will decay as the magnitude of the acceleration increases. This might have potential implications for black hole information problem. In this work, we address the physically more interesting question as to whether this entanglement can be retrieved in any manner. The most natural set-up for this is the quantum version of the twin paradox, in which we consider two initially inertial detectors, of which one undergoes a phase of acceleration and deceleration and returns back to its inertial phase, thus mimicking the quantum version of the twin paradox. We would like to study the individual probe response and entanglement profile of this system of twin quantum detectors. More specifically, we ask what happens to entanglement between the twin quantum detectors, initially uncorrelated, and coupled identically to an external quantum field. We identify novel features in the response of these detectors that can be traced to the change in the direction of acceleration. More importantly, we obtain the negativity measure for entanglement between these detectors and show that these exhibit some novel features. 

We use the standard setup, viz., a pair of two-level atomic probes $A$ and $B$ - the so-called Unruh-DeWitt detectors - weakly interacting with a background quantum scalar field $\Phi(x)$. The interaction Hamiltonian of the system is given by $
    \mathcal{H}_{\rm{int}}(\tau) = \sum_{I=A,B}\mu_I \,\chi(\tau_I)\,m_I(\tau_I)\,\Phi(x(\tau_I))~.
$
Here, $\mu$ is the interaction strength, $\chi(\tau)$ is the switching function, and $m(\tau)$ is the monopole operator. The detectors are initially uncorrelated, and thus the product state of the detector-field system in the asymptotic past can be expressed as, $\l|\rm{in}\r>=|0\rangle|g_A\rangle|g_B\rangle$, where $\l|0\r>$ is the vacuum state of the field. In the interaction picture, the state in the asymptotic future will be $\l|\rm{out}\r>=\mathcal{T}\l\{\exp{(i\int H_{\rm{int}}\,\DM \tau} \l|\rm{in}\r>)\r\}$ with $\mathcal{T}$ representing time ordering. For identical detectors with same interaction strength $\mu$, a perturbative expression for the reduced density matrix can be obtained by tracing over the field:$\rho_{\text{AB}}={\rm Tr}_{\phi}\, \ket{{\rm out}}\bra{{\rm out}}$. Various components of this matrix then characterize single detector transition rates as well as entanglement induced between the detectors. Their computation is standard (see, for instance, Ref. \cite{Koga:2018the}) and the key quantities turn out to be
\begin{subequations}\label{eq:Ij-Ie-general}
\begin{eqnarray}\label{eq:Ij-general}
\mathcal{I}_{\text{A}} &=& 
\int_{-\infty}^{\infty} d\tau^\prime \int_{-\infty}^{\infty}d\tau \;
\chi(\tau) \chi(\tau^{\prime}) 
e^{i \omega (\tau - \tau^{\prime})}
G_{W}(x_{\text{A}}(\tau^{\prime}), x_{\text{A}}(\tau))
\\ 
i \, \mathcal{I}_{\varepsilon} &=& \int_{-\infty}^{\infty} ds \int_{-\infty}^{\infty}d\tau \; \chi(\tau) \chi(s) 
e^{i \omega (\tau + s)}
G_{F}(x_{\text{B}}(s), x_{\text{A}}(\tau))~,
\label{eq:Ie-general}
\end{eqnarray}
\end{subequations}
where, $G_{W}(x^{\prime},x)$ and $G_F(x_2,x_1)$ respectively denote the Wightmann function and the Feynman propagators. As elaborated in \cite{Reznik:2002fz, Koga:2018the, Koga:2019fqh} and \cite{K:2023oon}, we define the measure of the entanglement through \textit{Negativity} $\mathcal{N}(\rho_{\text{AB}})$ whose expression in terms of above quantities is 
\begin{eqnarray}\label{eq:neg}
\mathcal{N}(\rho_{\text{AB}}) &=& \l[\sqrt{(\mathcal{I}_{\text{A}}-\mathcal{I}_{\text{B}})^2+4\,|\mathcal{I}_{\epsilon}|^2} - (\mathcal{I}_{\text{A}}+\mathcal{I}_{\text{B}})\r]/2 + \mathcal{O}(\mu^4) ~, 
\end{eqnarray}
We shall also study the \textit{mutual information} corresponding to shared correlations between the detectors, defined by
\begin{align}
    \mathcal{I}_{\text{AB}} = \mathcal{I}_{+} \ln \mathcal{I}_{+} + \mathcal{I}_{-} \ln \mathcal{I}_{-} - \mathcal{I}_{\text{A}}\ln \mathcal{I}_{\text{A}} - \mathcal{I}_{\text{B}} \ln \mathcal{I}_{\text{B}} + \mathcal{O}(\mu^4)
\end{align}
where $\mathcal{I}_{\pm}$ are defined as,
\begin{align}
    \mathcal{I}_{\pm} = \frac{1}{2} \l( \mathcal{I}_{\text{A}} + \mathcal{I}_{\text{B}} \pm \sqrt{\l(\mathcal{I}_{\text{A}} - \mathcal{I}_{\text{B}}\r)^2 + 4|\mathcal{I}_{AB}|^2 } \r)~.
\end{align}

The manuscript is organized as follows: In Sec. \ref{sec:Coords} we describe the kinematic set-up for the trajectories of the twins, focussing specifically on the nature of \textit{geodesics connecting points of simultaneity between respective trajectories.} This interval can generically turn from spacelike to timelike (if the twins initially sync their clocks). Section \ref{sec:Single-detector}
deals with the analysis of vacuum fluctuations as probed by the accelerating twin. Non-trivial effects in the transition probability of the accelerated observer due to non-uniform acceleration is obtained here. In Sec. \ref{sec:Entanglement}, we discuss the entanglement dynamics between the inertial and accelerated detectors by evaluating the negativity and mutual information shared between them. Finally, we conclude with a discussion and implications of our results in \ref{sec:discussions}.


\section{Twin paradox and geodesic separation}\label{sec:Coords}

In this section, we describe the trajectories for the twins. We also estimate the geodesic separation between these trajectories, which is essential for comprehending the entanglement dynamics between the detectors and relating it with their causal structure.

\subsection{Setting up the twins}

\begin{figure}[h!]
\centering
\includegraphics[width=0.45\textwidth]{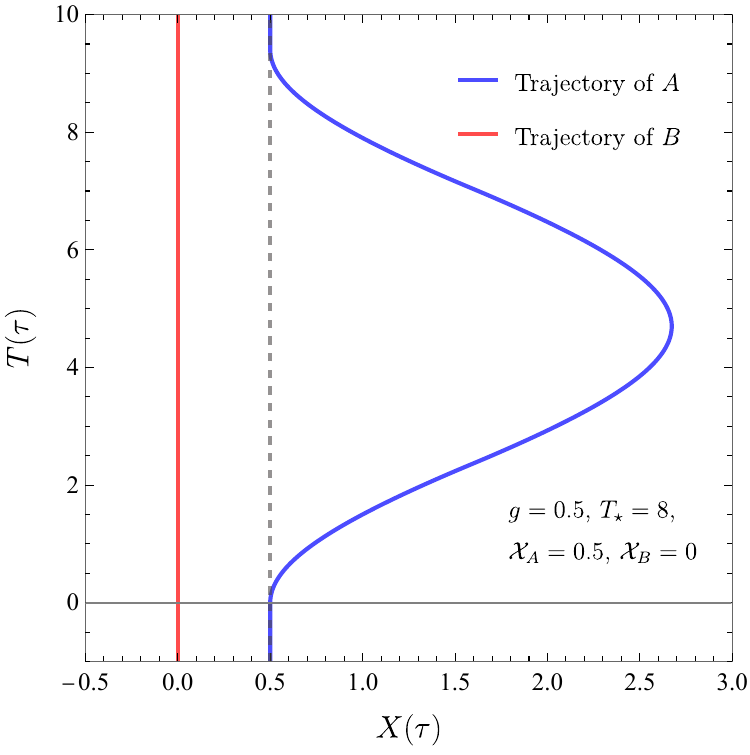}
\hskip 10pt
\includegraphics[width=0.45\textwidth]{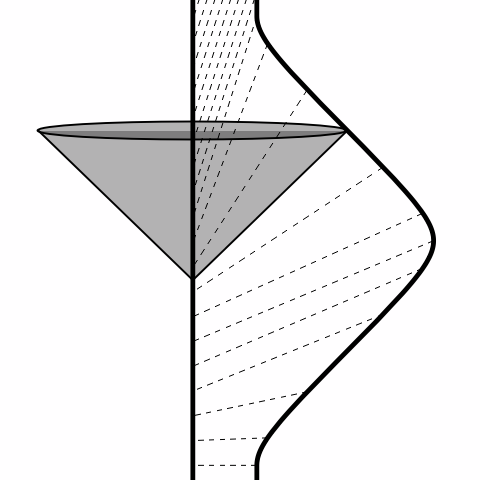}
\caption{In the above figure we have depicted the trajectories followed by the two twins Alice $(A)$ and Bob $(B)$. Initially both observers are separated by a spatial distance $X_0$. At time $T=0$, Alice undergoes a sequence of accelerated motions and at $T=T_{\star}$ returns to her inertial state at distance $x_0$ from Bob. The clocks of Alice and Bob are synced at $T=0$. The peculiarities of time dilation are highlighted in the structure of geodesics connecting instants of \textit{same values of proper times} on the two trajectories. These geodesics turn from spacelike to timelike, becoming null in between; see Fig. \ref{fig:Sigma2-sync}.}
\label{fig:TP-Trajectory}
\end{figure}

We consider the twins, Alice and Bob, denoted by $A$ and $B$. Initially, both twins are at rest and separated by a certain spatial distance which we will denote by $\mathcal{X}_{\text{A}}$ by choosing the rest frame of Bob. In particular, we consider $A$ to follow a trajectory of non-uniform acceleration. The trajectory of observer $A$ is described by
\begin{subequations}\label{eq:Trajec-TA&XA}
\begin{eqnarray}\label{eq:Trajec-TA}
    T_{A}(\tau) &=& T_{A}(\tau_{\star}) + \int_{\tau_{\star}}^{\tau} d\alpha\,\cosh{\kappa(\alpha)}~,\\
    X_{A}(\tau) &=& X_{A}(\tau_{\star}) + \int_{\tau_{\star}}^{\tau} d\alpha\,\sinh{\kappa(\alpha)}~.\label{eq:Trajec-XA}
\end{eqnarray}
\end{subequations}
In the previous expression, $T$ and $X$ respectively denote the Minkowski time and spatial coordinates, and $\tau$ denotes the proper time of the observer. $\tau_{\star}$ is some pivotal value of $\tau$ which set the initial conditions. For observer $A$, the quantity $\kappa(\alpha)$, see \cite{Minguzzi:2004fa}, is given by
\begin{equation}\label{eq:TA-Acc-Kpp}
    \kappa(\tau) = \int_{0}^{\tau}g(\alpha)\,d\alpha = 
    \begin{cases}
    g\,\tau &\tau\in \big[0,(T_{\star}/4)\big]\\
    -g\,\tau+g\,T_{\star}/2 &\tau\in \big[(T_{\star}/4),(3T_{\star}/4)\big]\\
    g\,\tau-g\,T_{\star} & \tau\in \big[(3T_{\star}/4),T_{\star}\big]~.
    \end{cases}
\end{equation}
where, $g$ is the constant acceleration and $T_{\star}$ is the total time duration of the non-zero acceleration. One can obtain the exact expressions for $T_{A}(\tau)$ and $X_{A}(\tau)$ as piecewise functions expressed in Eq. \eqref{eq:Trajec-TA&TB2} of Appendix \ref{Appn:Trajec-P1}. At the same time, the trajectory for Bob can also be obtained from Eq. \eqref{eq:Trajec-TA&XA} with the realization of $\kappa(\alpha) = 0$, as $B$ is a stationary observer. The exact representation of this trajectory is $T_{B}(\tau) = \tau_{B},\;X_{B}(\tau) = \mathcal{X}_{B}$.
In Fig. \ref{fig:TP-Trajectory}, we have depicted the trajectories for $A$ and $B$ for specific parameter values.

\subsection{Geodesic interval and Lines of simultaneity}

\begin{figure*}
\centering
\includegraphics[width=5.9cm]{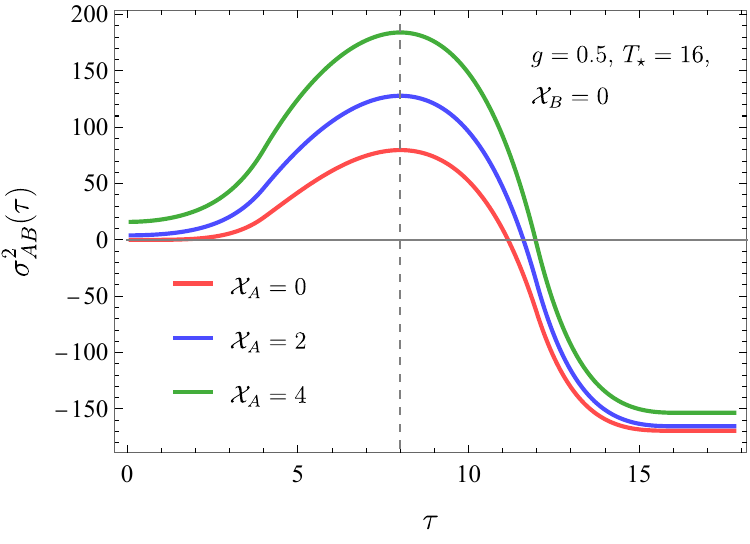}
\hskip 5 pt
\includegraphics[width=5.7cm]{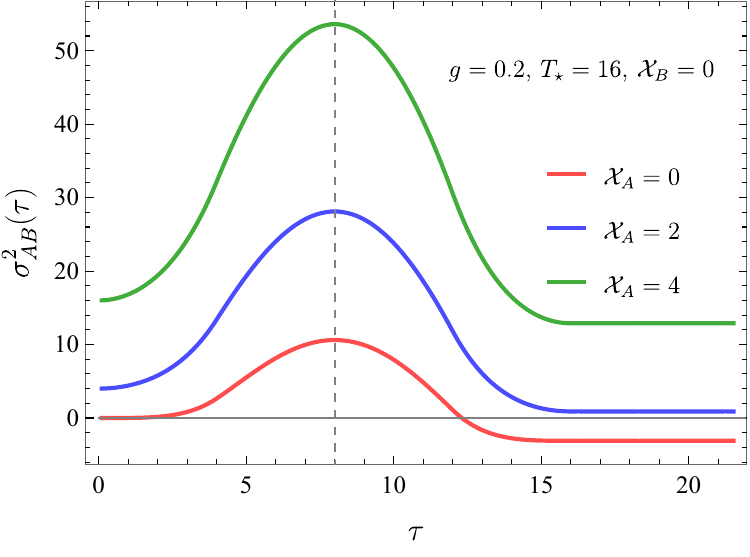}
\hskip 5 pt
\includegraphics[width=5.7cm]{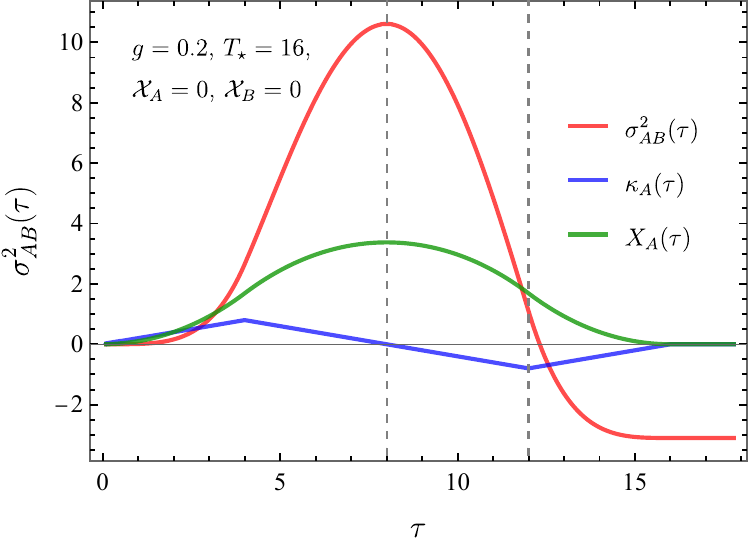}
\caption{The geodesic separation syncing the clocks of the twins Alice $A$ and  $B$, which is $\sigma^2_{A,B}(\tau,\tau)$ is plotted as the proper time $\tau$ progresses. One can notice that this quantity changes sign and becomes negative from positive as $\tau$ increases. Here $\mathcal{X}_{\text{A}}=X^{A}(\tau)$.}
\label{fig:Sigma2-sync}
\end{figure*}

To study the entanglement dynamics between the twins, one needs to calculate the two-point correlation function of the quantum field. At the same time, the key quantity in the two-point correlation function is the geodesic interval between the two observers. In Minkowski spacetime, one can easily compute this geodesic interval with the help of the expression $\sigma^2(x,x')=-(T-T')^2 + (X-X')^2$, and for our setup, it can be expressed as
\begin{eqnarray}\label{eq:sigma2-AB-TDA}
    \sigma^2_{A,B}(\tau_{A},\tau_{B}) &=& -\Big(T_{A}(\tau_{A})-T_{B}(\tau_{B})\Big)^2 + \Big(X_{A}(\tau_{A})-X_{B}(\tau_{B})\Big)^2~\nonumber\\
    &=& -\bigg(\int^{\tau_{A}}_{0}e^{\kappa(\alpha)}\,d\alpha+\mathcal{X}_{A}-\tau_{B}-\mathcal{X}_{B}\bigg)~\bigg(\int^{\tau_{A}}_{0}e^{-\kappa(\alpha)}\,d\alpha-\mathcal{X}_{A}-\tau_{B}+\mathcal{X}_{B}\bigg)~.
\end{eqnarray}
In the above expression, the quantity $\kappa(\alpha)$ should be considered from Eq. \eqref{eq:TA-Acc-Kpp}. In the special scenario of uniform acceleration, i.e., when $\kappa(\alpha)=g\,\alpha$ with $g$ being the acceleration of the detector, one can obtain the geodesic separation to be
\begin{eqnarray}\label{eq:sigma2-AB-TIA}
    \sigma^2_{A,B} 
    &=& -\bigg\{\frac{1}{g}\big(e^{g\,\tau_{A}}-1\big)+\mathcal{X}_{A}-\tau_{B}-\mathcal{X}_{B}\bigg\}~\bigg\{-\frac{1}{g}\big(e^{-g\,\tau_{A}}-1\big)-\mathcal{X}_{A}-\tau_{B}+\mathcal{X}_{B}\bigg\}~.
\end{eqnarray}
In Fig. \ref{fig:Sigma2-sync}, we have plotted the geodesic interval of Eq. \eqref{eq:sigma2-AB-TDA} when clocks associated with both the probes are synchronized, i.e., when $\tau_{A}=\tau_{B}$.

\section{Excitations due to non-uniform acceleration}\label{sec:Single-detector}

\begin{figure*}
\centering
\includegraphics[width=7.8cm]{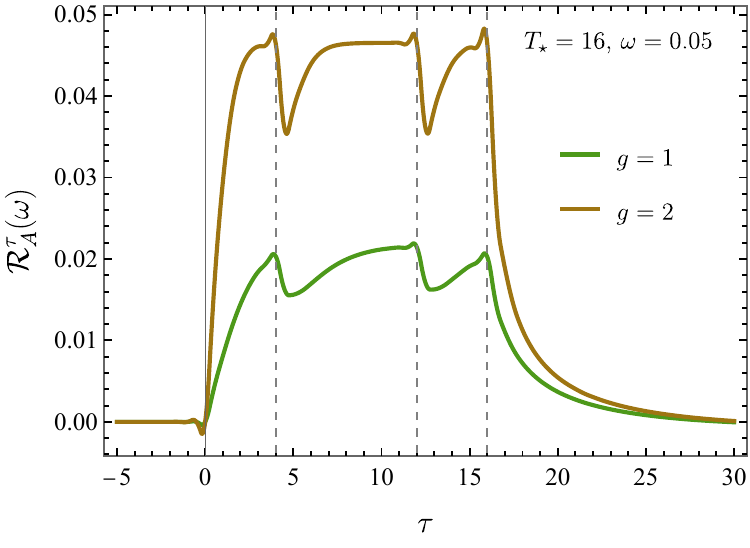}
\hskip 30pt
\includegraphics[width=7.8cm]{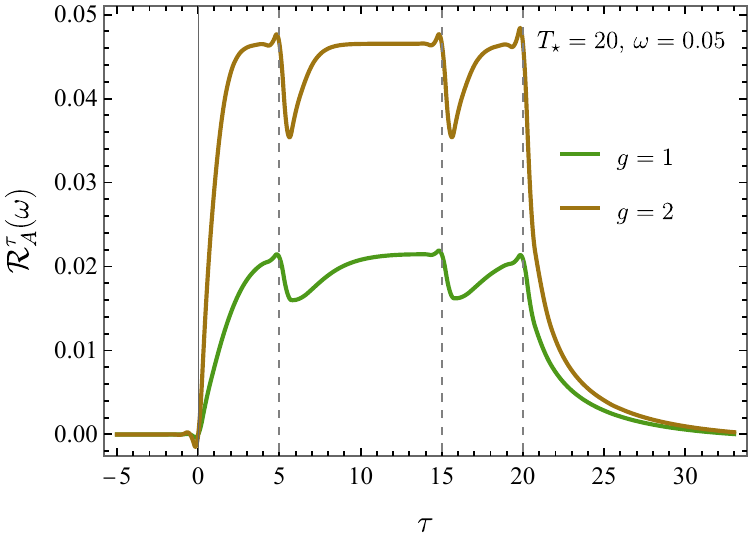}
\caption{Single detector transition probability rates $R_{A}(\omega)$ are plotted as functions of the detection time $\tau$. On the left, we have considered fixed $T_\star=16$, and on right we have considered $T_\star=20$. In both cases we have fixed $g=2$. From  these plots one can observe that there are certain peaks and dips whenever the probe $A$ with non-uniform acceleration changes the direction of its acceleration.}
\label{fig:IA-NonUni-Acc}
\end{figure*}
In this section, our main goal is to obtain the excitation probability rate for a quantum two-level atom due to the vacuum fluctuations sampled by the non-uniform accelerated motion. Before getting into that, we recall the excitation probability rate for an atom in inertial motion, which corresponds to the trajectory of probe $B$. In the case of inertial probe $B$, the geodesic interval $\sigma^2_{B,B}(\tau_{B},\tau'_{B})=-(\tau_{B}-\tau'_{B})^2$. The Wightman function is obtained as
\begin{eqnarray}\label{eq:W-BB}
    \mathcal{W}(\tau_{B},\tau'_{B}) = -\frac{1}{4\,\pi^2}\,\frac{1}{(\tau_{B}-\tau'_{B}-i\,\epsilon)^2}~.
\end{eqnarray}
With the help of this Wightman function, one can get the transition probability rate utilising Eq. \eqref{eq:Ij-general}. In particular, the transition probability rate is defined as $\mathcal{R}^{\tau}_{j} = \mathcal{I}_{j}/(\mu^2\,\mathcal{T})$ with $\mathcal{T} = \lim_{T\to\infty}(1/2)\int_{-T}^{T}ds$ for eternal switching. For probe $B$ this transition probability rate is
\begin{eqnarray}\label{eq:IBD-omega}
    \mathcal{R}_{B}^{\tau}(\omega) &=& -\frac{1}{4\,\pi^2}\int_{0}^{\infty}ds\,\bigg[\frac{e^{-i\,\omega\,s}}{(s-i\,\epsilon)^2}+\frac{e^{i\,\omega\,s}}{(s+i\,\epsilon)^2}\bigg]\nonumber\\
    ~&=& -\frac{1}{4\,\pi^2}\int_{-\infty}^{\infty}ds\,\frac{e^{-i\,\omega\,s}}{(s-i\,\epsilon)^2}~.
\end{eqnarray}
For $\omega>0$ the above integration should be carried out considering a contour in the lower half complex plane. However, the pole in the upper half complex plane, which implies that the above integral should vanish. Therefore, the excitation rate for detector $B$ vanishes, which is consistent with the general understanding from the existing literature \cite{Barman:2023aqk, Barman:2025lsq}.\vspace{0.2cm}

Next, we consider the important case -  non-uniform acceleration of probe $A$ - which follows a trajectory given by Eqs. \eqref{eq:Trajec-TA&XA} and \eqref{eq:TA-Acc-Kpp}. To obtain the transition probability rate of this probe, we consider the prescription as mentioned in \cite{Schlicht:2003iy}. In the expression of the transition probability rate, the Wightman function corresponding to probe $A$ can be expressed as
\begin{eqnarray}\label{eq:Wtmn-Fn-PrbA1}
    \mathcal{W}(\tau,\tau') = -\frac{1}{4\,\pi^2}\,\frac{1}{\mathcal{I}_{+}(\tau,\tau')\times \mathcal{I}_{-}(\tau,\tau')}~,
\end{eqnarray}
where 
\begin{eqnarray}\label{eq:Ipm-PrbA1}
    \mathcal{I}_{\pm}(\tau,\tau') &=& \int_{\tau}^{\tau'}d\alpha\,\exp{\l\{\kappa(\alpha)\r\}}~.
\end{eqnarray}
We would like to mention that the above representation was previously utilised to obtain the transition probability of probes in non-uniform acceleration in \cite{Kothawala:2009aj}. In that work, the authors could analytically obtain the transition probability rates considering perturbative corrections to the non-uniform acceleration profile as compared to a uniform acceleration. In the current scenario, we want to provide the transition probability rates without these kinds of perturbative considerations. We observed that it is not possible to obtain analytical expressions, at least to our knowledge, for $\mathcal{I}_{\pm}(\tau,\tau')$ utilizing the acceleration profiles of Eq. \eqref{eq:TA-Acc-Kpp}. However, one can obtain $\mathcal{I}_{\pm}(\tau,\tau')$ using numerical methods and then utilize them to numerically obtain the transition probability rates. Here, we would also like to mention that to obtain this transition probability rate, we have subtracted $\big\{-e^{-i\,\omega\,s}/(4\pi^2\,s^2)\big\}$ from the integrand of Eq. \eqref{eq:Ij-general}, as a form of regularization. It should be noted that the above subtracted quantity corresponds to the transition probability rate of an inertial probe. The resulting transition probability rate is plotted in Fig. \ref{fig:IA-NonUni-Acc}, which corresponds to the eternal switching scenario. We must mention here that one obtains results with same features with a Gaussian switching function using numerical integration\footnote{While this work was being written up, a preprint with similar analysis appeared on arXiv \cite{Zambianco:2025ehv}; it might be interesting to compare their results with our analysis here. In this context, see also \cite{Dubey:2025wws}.}.

\section{Entanglement between twins} \label{sec:Entanglement}
For the generation of non-local correlation-entanglement-between the quantum probes through the interaction of the quantum field requires one to compute the two-point correlation of the quantum field. This implies that any non-local correlation acquired by the quantum probes that are initially uncorrelated, is due to the `harvesting' of correlations (entanglement) from the quantum field. As mentioned in the introduction, some entanglement measures that are helpful to quantify the entanglement harvested by the quantum probes include entanglement negativity, mutual information etc. Any such quantities that quantify the correlations between the quantum probes, involve the interplay between local noise and non-local correlations between the detectors. We first analyze the entanglement measure negativity in the context of twin paradox and then evaluates mutual information to quantify any correlations harvested between the twins due to their differential time  

\subsection{Entanglement negativity} \label{sec:negativity}
\begin{figure}
    \centering
    \includegraphics[width=0.48\linewidth]{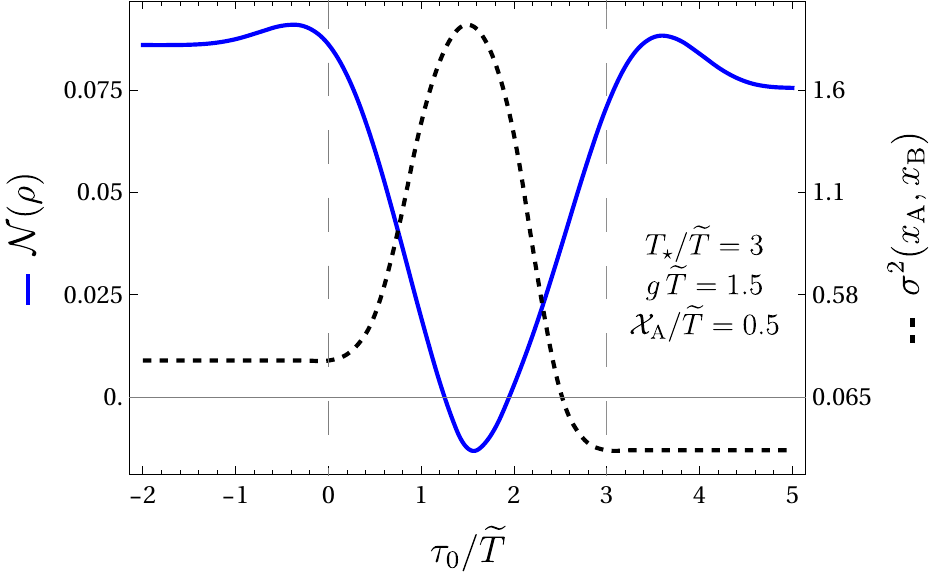}
    \hskip 5pt
    \includegraphics[width=0.48\linewidth]{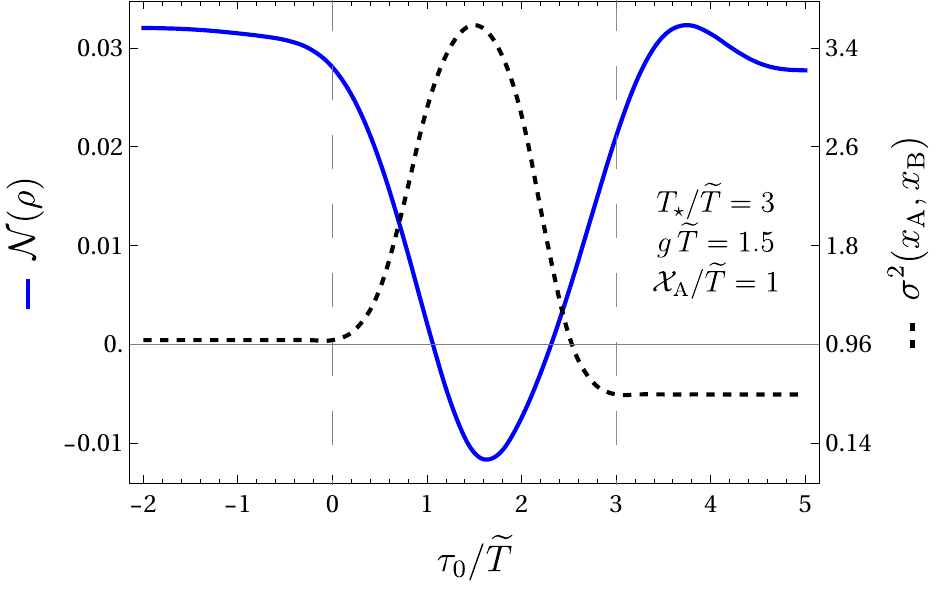}
    \vskip 5pt
    \includegraphics[width=0.48\linewidth]{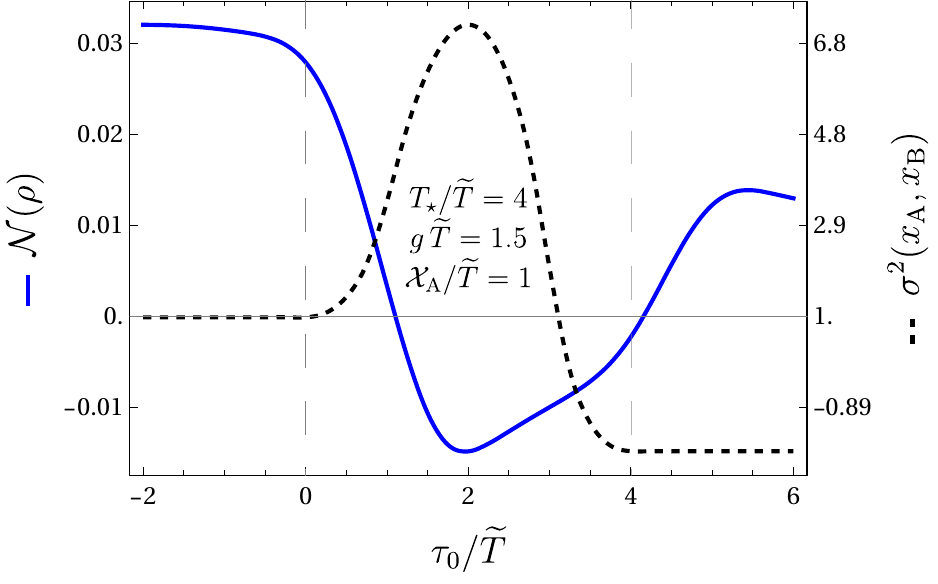}
    \hskip 5pt
    \includegraphics[width=0.48\linewidth]{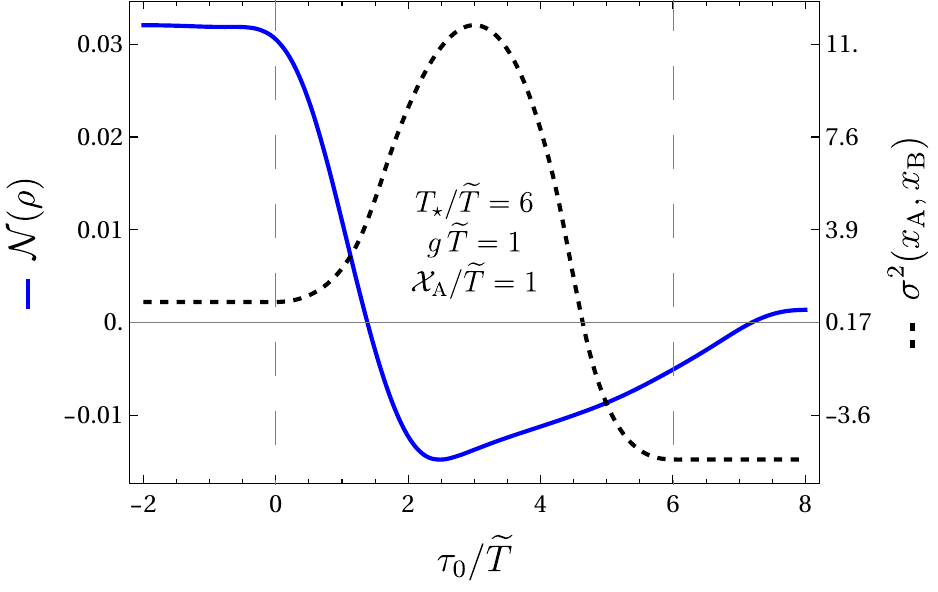}
    \caption{Negativity $\mathcal{N}$ at different times of simultaneity. The dashed curve indicate the geodesic interval along the lines of simultaneity - i.e it refers to the separation measured using geodesics connecting same values of proper time. The scaled energy gap, $\omega/\widetilde{T}$ is chosen as unity for all the plots.} 
    \label{fig:Neg-vs-tau0}
\end{figure}

The negativity is estimated using the individual detector transitions and non-local correlations between the detectors. The individual detector transitions characterizes the vacuum fluctuations of the field in the local region of the individual observers whereas non-local correlations characterizes the genuine entanglement manifested from the quantum field. In Sec. \ref{sec:Single-detector}, we have already arrived at the single detector transition for both the twins. To estimate the non-local correlation, $\mathcal{I}_{\mathcal{E}}$, consider the Eq. \ref{eq:Ie-general}. The Feynman propagated useful to estimate $\mathcal{I}_{\mathcal{E}}$ is given by,
\begin{align}
    G_{F}(x,x')=\l(\frac{i}{4\pi^2}\r) \frac{1}{-(T_{\text{A}}(\tau)-T_{\text{B}}(\tau'))^2 + (X_{\text{A}}(\tau)-X_{\text{B}}(\tau'))^2 + i \epsilon}~.
\end{align}
Using the form of spacetime form of Feynman propagator in Eq. \ref{eq:Ie-general} gives, 
\begin{align}
    \mathcal{I}_{\mathcal{E}}= \int_{-\infty}^{\infty} \DM\tau \int_{-\infty}^{\infty} \DM \tau' \l(\frac{i}{4\pi^2}\r)\frac{e^{(\tau-\tau_0)^2/2\widetilde{T}} \, e^{(\tau'-\tau_0)^2/2\widetilde{T}}\, e^{i\omega (\tau+\tau')}}{-(T_{\text{A}}(\tau)-T_{\text{B}}(\tau'))^2 + (X_{\text{A}}(\tau)-X_{\text{B}}(\tau'))^2 + i \epsilon}
\end{align}
where we have use the Gaussian switching function to control the interaction between the quantum field and the quantum probe. One can express the Gaussian function as an integral, $e^{(\tau-\tau_0)^2/2\widetilde{T}} = \int_{-\infty}^{\infty} \kappa\,(\widetilde{T}/\sqrt{2}) e^{-\kappa ^2\widetilde{T}^2/2} e^{-i\kappa (\tau - \tau_0)} $ and then use contour integration to evaluate the integral. The relevant details of the calculations are sketched in the Appendix \ref{app:cross-correlations}. With individual detector transitions and cross correlation we can now evaluate the negativity that corresponds to the entanglement acquired by the quantum probes carried by the twins using the relation,
\begin{eqnarray}\label{eq:neg}
\mathcal{N}(\rho_{\text{AB}}) &=& \l[\sqrt{(\mathcal{I}_{\text{A}}-\mathcal{I}_{\text{B}})^2+4\,|\mathcal{I}_{\epsilon}|^2} - (\mathcal{I}_{\text{A}}+\mathcal{I}_{\text{B}})\r]/2 + \mathcal{O}(\mu^4) ~,
\end{eqnarray}
where $\mathcal{I}_{\text{A,B}}$ are the individual detector transitions and $\mathcal{I}_{\mathcal{E}}$ is the cross correlation. There are different length scales in our setup - switching time scale $\widetilde{T}$, non-uniform acceleration time scale $T_{\star}$, acceleration length scale $1/g$, energy gap $\omega$, initial spatial separation between the twins $\mathcal{X}_{\text{A}}$ and proper time at the peak of switching $\tau_0$. To understand the effect of differential aging between the twins in the negativity, our interest will be focused on relationship between negativity and peak of the switching function as it varies from inertial to non-uniform acceleration and back to inertial. The values of other parameters are suitably chosen such that we consider the situation where entanglement is harvested between the detectors. Fig. \ref{fig:Neg-vs-tau0} depicts how different parameters can affect negativity. It also illustrates how negativity initially decreases and then increases with time.

Initially, when the proper time of the twins are synchronized, they start their journey at the same time. At this instant, the geodesic interval between the twins is spacelike. As one of them undergo non-uniform acceleration, due to differential flow of time, the geodesic interval between them can become timelike depending on the magnitude of acceleration and time scale of acceleration. This change in the causal structure between the two observers due to acceleration has its imprint on negativity. 

The negativity measure for the detectors shows an increase in after the acceleration stage has completed. To clarify this, we consider the non-local correlation that should be associated with the actual entanglement between the detectors. From the definition of negativity that we have used, one has to explicitly isolate the non-local correlation to understand how non-uniform acceleration can play an important role in entanglement. We follow the same procedure as our previous work \cite{K:2023oon} (see also \cite{Ng:2018ilp}) for this purpose, and define
\begin{align}
    \mathcal{N}^{+} = \l[\sqrt{(\mathcal{I}_{\text{A}}-\mathcal{I}_{\text{B}})^2+4\,|\mathcal{I}^{+}_{\epsilon}|^2} - (\mathcal{I}_{\text{A}}+\mathcal{I}_{\text{B}})\r]/2
\end{align}
to obtain the dominant contribution from the genuine entanglement.

\begin{figure}
    \centering
    \includegraphics[width=0.48\linewidth]{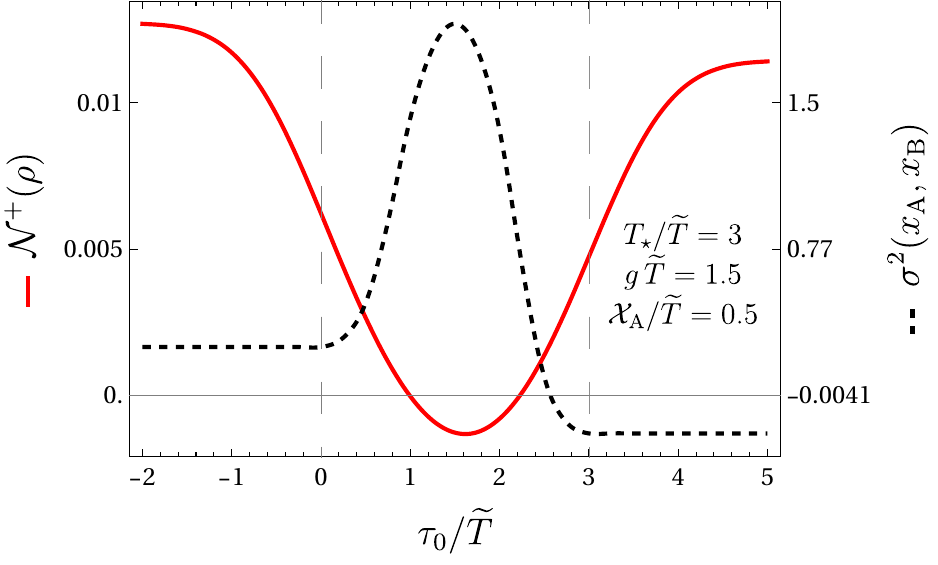}
    \hskip 5pt
    \includegraphics[width=0.48\linewidth]{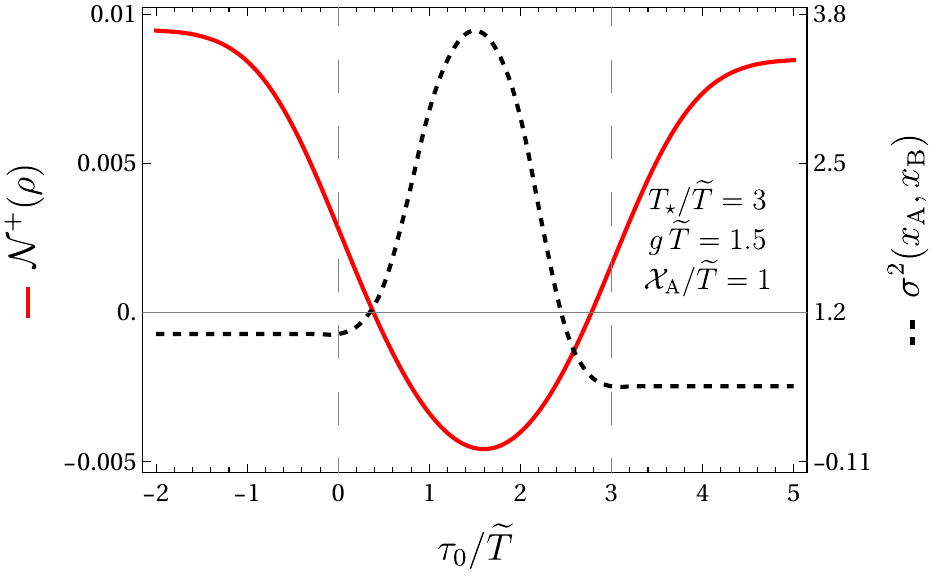}
    \vskip 5pt
    \includegraphics[width=0.48\linewidth]{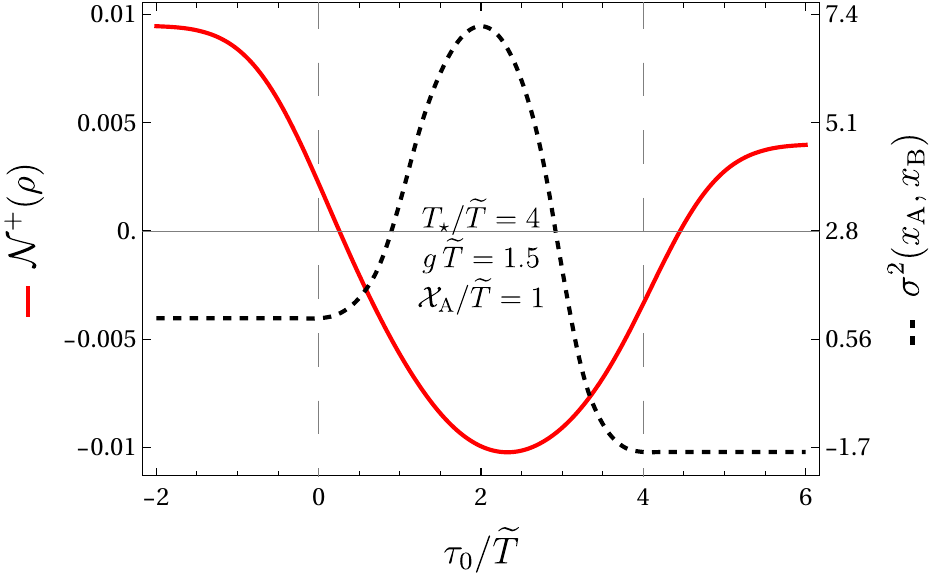}
    \hskip 5pt
    \includegraphics[width=0.48\linewidth]{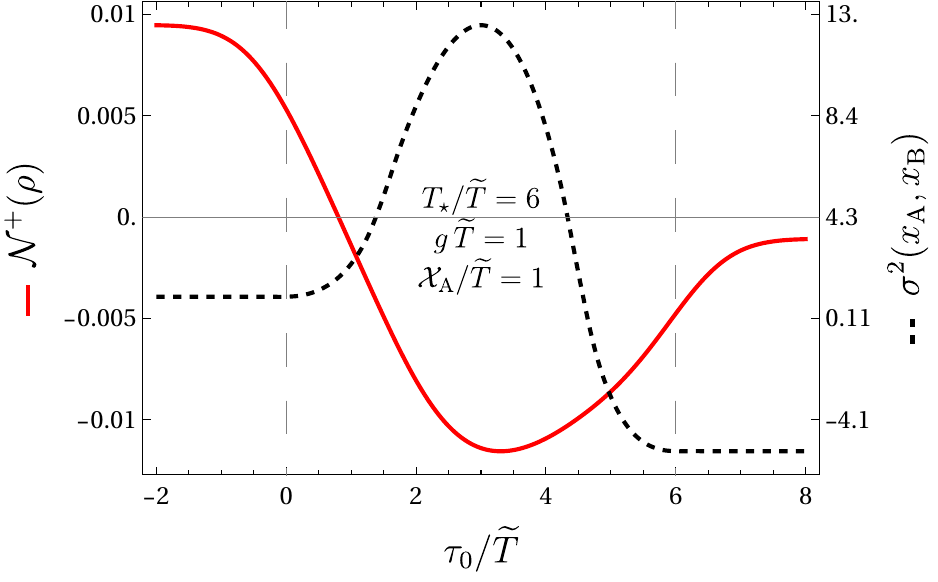}
    \caption{The non-local part of Negativity $\mathcal{N}^{+}$ that quantify the genuine entanglement at different times of simultaneity. The dashed curve indicate the geodesic interval along the times of simultaneity. The genuine entanglement retrieved is saturates before reaching the initial value of negativity. The solid horizontal line indicate the zero of the negativity. Counting from clockwise, all three except the second scenario, the geodesics interval changes from spacelike to timelike. The scaled energy gap, $\omega/\widetilde{T}$ is chosen as unity for all the plots.}
    \label{fig:Neg-plus-vs-tau0}
\end{figure}

\subsection{Mutual Information} \label{sec:mutinf}
\begin{figure}
    \centering
    \includegraphics[width=0.45\linewidth]{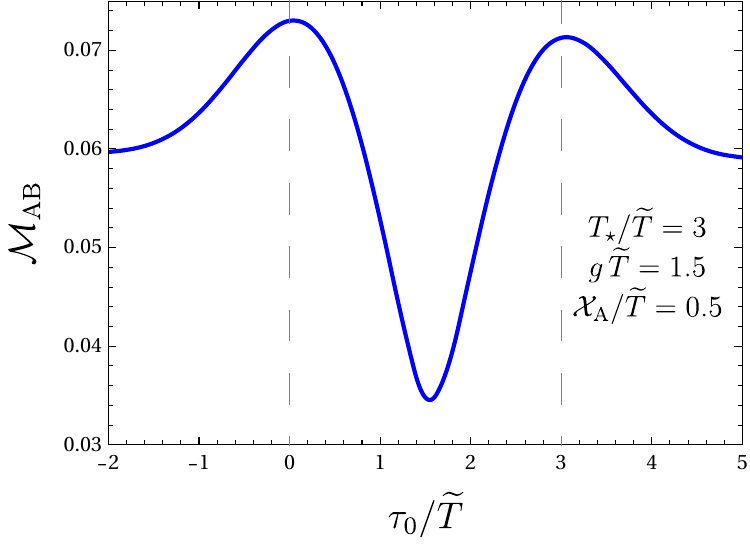}
    \hskip 15pt
    \includegraphics[width=0.45\linewidth]{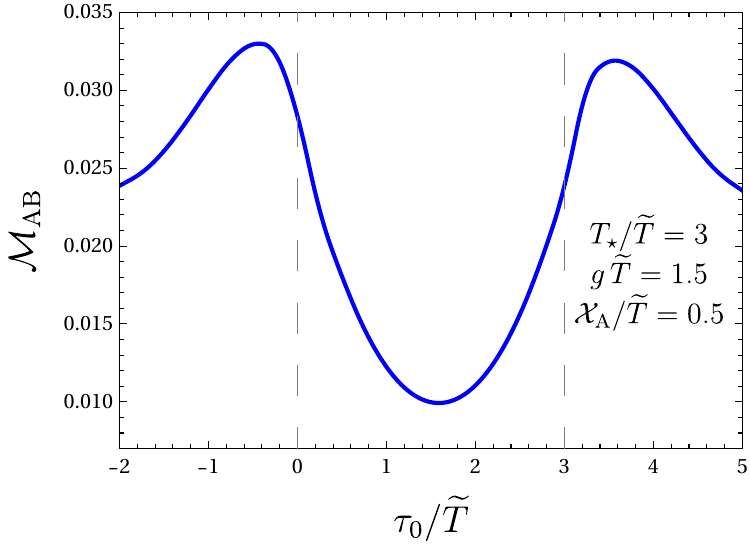}
    \vskip 5pt
    \includegraphics[width=0.45\linewidth]{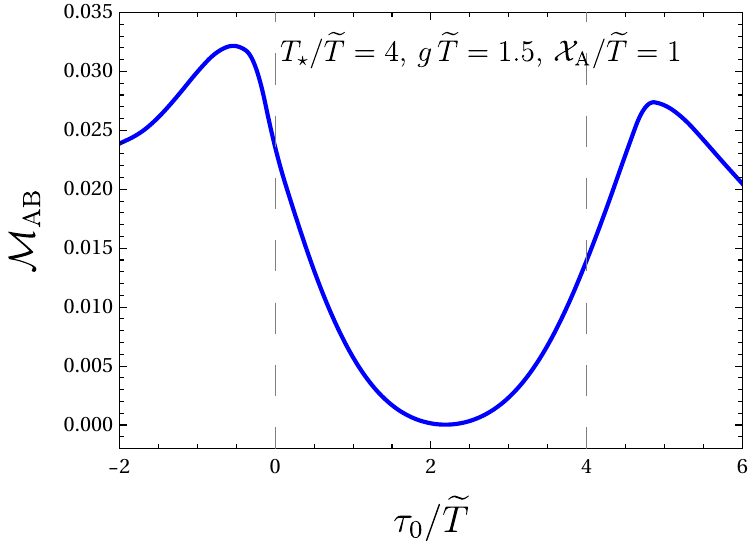}
    \hskip 15pt
    \includegraphics[width=0.45\linewidth]{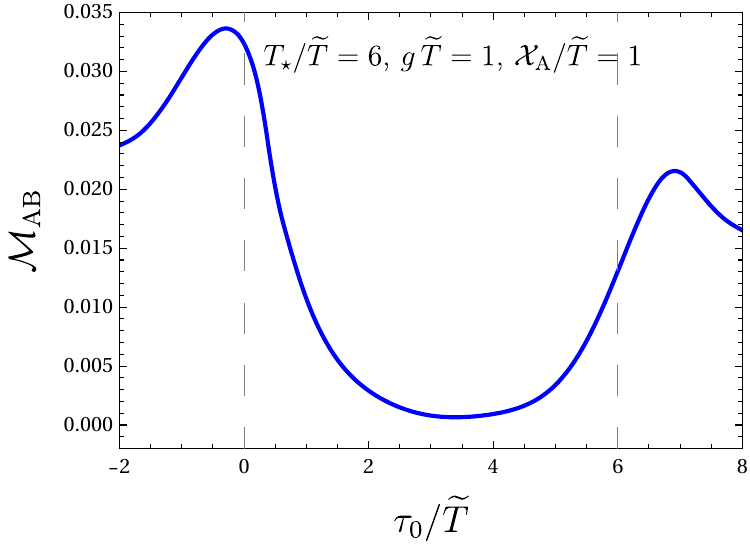}
    \caption{The behaviour of mutual information $\mathcal{M}_{AB}$ at times of simultaneity. The disturbance of the field - captured by the field correlations - as the twin starts starts and stops accelerating is evident. The change in direction of acceleration increases the facilitates the \textit{retrieval} the lost correlations between the quantum systems due to the coupling with quantum field. The scaled energy gap, $\omega/\widetilde{T}$ is chosen as unity for all the plots.}
    \label{fig:MI-vs-tau0}
\end{figure}

Mutual information measures the correlation between the two quantum subsystems. In our context, these subsystems are the 2-level atomic detectors carried by Alice and Bob and in this case the mutual information can be determined using the relation, 
\begin{align}
    \mathcal{M}_{\text{AB}} = \mathcal{I}_{+} \ln \mathcal{I}_{+} + \mathcal{I}_{-} \ln \mathcal{I}_{-} - \mathcal{I}_{\text{A}}\ln \mathcal{I}_{\text{A}} - \mathcal{I}_{\text{B}} \ln \mathcal{I}_{\text{B}} + \mathcal{O}(\mu^4)
\end{align}
where $\mathcal{I}_{\pm}$ are defined as,
\begin{align}
    \mathcal{I}_{\pm} = \frac{1}{2} \l( \mathcal{I}_{\text{A}} + \mathcal{I}_{\text{B}} \pm \sqrt{\l(\mathcal{I}_{\text{A}} - \mathcal{I}_{\text{B}}\r)^2 + 4|\mathcal{I}_{AB}|^2 } \r)~.
\end{align}
where $\mathcal{I}_\text{A}$ and $\mathcal{I}_{\text{B}}$ are the individual transition probabilities and $\mathcal{I}_{\text{AB}}$ corresponds to the cross transition probabilities between detectors carried by A and B. The cross transition probability can be evaluated using,
\begin{align}
    \mathcal{I}_{\text{AB}} = \int_{-\infty}^{\infty} \DM \tau \int_{-\infty}^{\infty} \DM \tau' \,\chi(\tau) \chi(\tau') e^{-i\omega(\tau -\tau')}G^{+}(x_{\text{A}}^i(\tau),x_{\text{B}}^{i}(\tau'))
\end{align} 
Using the Gaussian switching function and form of Wightman function this can written as,
\begin{align}
    \mathcal{I}_{\text{AB}} = \int_{-\infty}^{\infty} \DM \tau \int_{-\infty}^{\infty} \DM \tau' \,\l(\frac{-1}{4\pi^2}\r)\frac{e^{(\tau-\tau_0)^2/2\widetilde{T}} \, e^{(\tau'-\tau_0)^2/2\widetilde{T}}\, e^{i\omega (\tau-\tau')}}{-(T_{\text{A}}(\tau)-T_{\text{B}}(\tau') -i \epsilon)^2 + (X_{\text{A}}(\tau)-X_{\text{B}}(\tau'))^2}
\end{align}
where we have used the same definitions as earlier. We follow the same steps as before by using the Fourier transform of the Gaussian switching function and then performing contour integration. Evaluating the residue at the poles and closing the contour in lower-half plane we compute the integral with respect to $\tau'$. The integration with respect to $\tau$ is done numerically. The behavior of mutual information is same as negativity and is illustrated in Fig. \ref{fig:MI-vs-tau0}. The change in causal structure due to the non-uniform acceleration is imprinted in the mutual information. The mutual information may not allow one to extract the genuine non-local correlations, hence it will also have contributions from local corelation between the detectors. Hence the behavior of mutual information is close(or not) to entanglement negativity. The initial acceleration, as expected will lead to decay in the mutual information \cite{FuentesSchuller:2004xp} and the change in acceleration leads to retrieving back correlations between the detectors carried by the twin observers.  

\section{Results and Conclusions} \label{sec:discussions}

The entanglement between quantum systems in inertial (Alice) and non-uniform accelerated (Bob) trajectories is sensitive to magnitude of acceleration and time duration of acceleration. The twin observers Alice and Bob starts their journey by synchronizing their clock. From Fig. \ref{fig:Neg-plus-vs-tau0} and \ref{fig:Neg-vs-tau0} it is evident that the entanglement negativity between the detectors carried by the twins decreases with initial acceleration when the Alice moves away from Bob. The decrease in negativity can be associated with the sudden increase in the geodesic interval between the observers as the correlations drop as $1/\sigma^2$. The change in direction of acceleration of Alice leads to retrieval of correlation between the quantum probes mediated by the quantum field. It is worth noting that the transition of the geodesic interval (connecting points with same proper times) from spacelike to timelike leaves no imprint on the entanglement.

Quantum probes localized on specific trajectories offer unique possibilities to probe spacetime over and above those offered by classical probes. This fact was specifically highlighted in recent works \cite{K:2023oon, Mayank:2025bkc} where the possibility of extracting curvature information from entanglement features of two such probes was analyzed and found to be feasible. In this work, we focus on specific trajectories in Minkowski spacetime characterizing the classical ``twin paradox", and replace the twins with quantum detectors coupled to an external quantum field. The entanglement thus induced carries imprints of the acceleration profile. Two key results of our work, characterizing entanglement between twin quantum systems, we derived are: (i) entanglement initially decreases and then increases as the twin approach each other, saturating to different values in asymptotic past and future, and (ii) change in direction of acceleration leaves imprints on the negativity profile. 

Our set-up and analysis also provide a novel way to probe the relation between quantum entanglement on the one hand, and closed timelike curves on the other, which in this setup can be obtained by connecting the twins through a wormhole. We intend to present results on this interesting aspect of the problem in a follow-up work.

\begin{acknowledgments}
K. H. would like to thank the Indian Institute of Technology Bombay (IIT Bombay) for supporting this work through a postdoctoral fellowship.
\end{acknowledgments}

\appendix

\section{Trajectory of non-uniform observer}\label{Appn:Trajec-P1}
In this section of the Appendix, we provide explicit expressions for the time and spatial coordinates that signify the coordinate transformation to a trajectory of non-uniform acceleration. These are obtained by integrating the expressions of Eq. \eqref{eq:Trajec-TA&XA} and are given by
\begin{subequations}\label{eq:Trajec-TA&TB2}
\begin{eqnarray}\label{eq:Trajec-TA2}
    T_{A} (\tau) &=&
    \begin{cases}
    \tau, & \tau \leq0  \\
    \dfrac{1}{g}\sinh(g\tau)
    & \tau \in (0, T_{\star}/4) \\
    \dfrac{1}{g}\sinh\l(g(\tau - T_{\star}/2)\r)
    + \dfrac{2}{g}\sinh\l(\dfrac{g T_{\star}}{4}\r)
    & \tau \in (T_{\star}/4, 3T_{\star}/4) \\
    \dfrac{1}{g}\sinh\l(g(\tau - T_{\star})\r)
    + \dfrac{4}{g}\sinh\l(\dfrac{g T_{\star}}{4}\r)
    & \tau \in (3T_{\star}/4, T_{\star}) \\
    \tau - T_{\star} + \dfrac{4}{g} \sinh \l(\dfrac{g T_{\star}}{4} \r) & \tau \geq T_{\star}~,
    \end{cases}\\
    X_{\text{A}}(\tau) &=&
    \begin{cases}
    \mathcal{X}_{A} & \tau \leq 0
    \\
    \mathcal{X}_{A} + \dfrac{1}{g}\cosh(g\tau) - \dfrac{1}{g}
    & \tau \in (0, T_{\star}/4) \\
    \mathcal{X}_{A} + \dfrac{1}{g}\cosh\l(g(\tau - T_{\star}/2)\r)
    + \dfrac{2}{g}\cosh\l(\dfrac{g T_{\star}}{4}\r) - \dfrac{1}{g}
    & \tau \in (T_{\star}/4, 3T_{\star}/4) \\
    \mathcal{X}_{A} + \dfrac{1}{g}\cosh \l(g(\tau - T_{\star})\r) - \dfrac{1}{g}
    & \tau \in (3T_{\star}/4, T_{\star})\\
    \mathcal{X}_{A}
    & \tau \geq T_{\star}~.
    \end{cases}\label{eq:Trajec-XA2}
\end{eqnarray}
\end{subequations}

\section{Estimating cross-correlation term $\mathcal{I}_{\mathcal{E}}$} \label{app:cross-correlations}

The integral for evaluating the cross correlation part is given by:
\begin{align}
    \mathcal{I}_{\mathcal{E}}= \int_{-\infty}^{\infty} \DM\tau \int_{-\infty}^{\infty} \DM \tau' \l(\frac{i}{4\pi^2}\r)\frac{e^{(\tau-\tau_0)^2/2\widetilde{T}} \, e^{(\tau'-\tau_0)^2/2\widetilde{T}}\, e^{i\omega (\tau+\tau')}}{-(T_{\text{A}}(\tau)-T_{\text{B}}(\tau'))^2 + (X_{\text{A}}(\tau)-X_{\text{B}}(\tau'))^2 + i \epsilon}
\end{align}
expressing the Gaussian switching function as $e^{(\tau'-\tau_0)^2/2\widetilde{T}} = \int_{-\infty}^{\infty} \DM \kappa \,(\widetilde{T}/\sqrt{2\pi}) e^{-\kappa^2\widetilde{T}^2/2} e^{-i\kappa(\tau' - \tau_0)} $ and the trajectory of the inertial observer,$T_{\text{B}}(\tau')=\tau',\; X_{\text{B}}(\tau')=\mathcal{X}_{\text{B}}$, the expression becomes,
\begin{align}
        \mathcal{I}_{\mathcal{E}}= \int_{-\infty}^{\infty}\DM \kappa\int_{-\infty}^{\infty} \DM\tau \int_{-\infty}^{\infty} \DM \tau'\frac{\widetilde{T}}{\sqrt{2\pi}} \l(\frac{i}{4\pi^2}\r)\frac{ e^{-\kappa^2\widetilde{T}^2/2} e^{-i\kappa(\tau' - \tau_0)} \, e^{(\tau-\tau_0)^2/2\widetilde{T}}\, e^{i\omega (\tau+\tau')}}{-(T_{\text{A}}(\tau)-\tau')^2 + (X_{\text{A}}(\tau)-\mathcal{X}_{\text{B}})^2 + i \epsilon}
\end{align}
We can do the contour integration for $\tau'$ and the poles are at $\tau'=T_{\text{A}}(\tau) \pm  \sqrt{(X_\text{A}(\tau) - X_B)^2 + i\,\epsilon}$. The poles are in both upper and lower plane and we find the residue of the integrand at these poles. Using these residues, and contour can be closed in upper half plane for $k\in [\omega,\infty)$ and in the lower half plane for $k \in (\infty,\omega]$. The last integral in $\tau$ is done numerically.



\end{document}